\documentclass[12pt,preprint]{aastex}

\usepackage{emulateapj5}
\usepackage{epsf}
\usepackage{graphics}

\newcommand{\kms}{km\,s$^{-1}$}
\newcommand{\dv}{$R^{1/4}\,$}

\newcommand{\mlsun}{M$_{\odot}$/L$_{\odot,B}$}

\newenvironment{inlinefigure}{
\def\@captype{figure}
\noindent\begin{minipage}{0.999\linewidth}\begin{center}}
{\end{center}\end{minipage}\smallskip}

\slugcomment{submitted to ApJ}

\shorttitle{The internal structure of MG2016+112}
\shortauthors{Treu \& Koopmans}

\begin{document}

\title{The internal structure and formation of early-type galaxies:
the gravitational--lens system MG2016+112 at
$z$=1.004}

\footnotetext[1]{Based on observations collected at W.~M. Keck
Observatory, which is operated jointly by the California Institute of
Technology and the University of California, and with the NASA/ESA
Hubble Space Telescope, obtained at STScI, which is operated by AURA,
under NASA contract NAS5-26555.}

\author{Tommaso Treu}
\affil{California Institute of Technology, 
Astronomy, mailcode 105--24, Pasadena, CA 91125}
\author{L\'eon V.E. Koopmans} 
\affil{California Institute of Technology, 
Theoretical Astrophysics Including Relativity (TAPIR), mailcode 
130--33, Pasadena, CA 91125}

\begin{abstract}
We combine our recent measurements of the velocity dispersion and the
surface brightness profile of the lens galaxy D in the system
MG2016+112 ($z=1.004$) with constraints from gravitational lensing to
study its internal mass distribution. We find that: (i) dark matter
accounts for $>$50\% of the total mass within the Einstein radius
(99\% CL), whereas $\sim$75\% is the more likely contribution. In
particular, we can exclude at the 8--$\sigma$ level that mass follows
light inside the Einstein radius with a constant mass-to-light ratio
(M/L). (ii) the total mass distribution inside the Einstein radius is
well-described by a density profile $\propto r^{-\gamma'}$ with an
effective slope $\gamma'=2.0\pm0.1\pm0.1$, including random and
systematic uncertainties. (iii) The offset of galaxy D from the local
Fundamental Plane independently constrains the stellar M/L, and
matches the range derived from our models, leading to a more stringent
lower limit of $>$60\% on the fraction of dark matter within the
Einstein radius (99\%~CL).

Under the assumption of adiabatic contraction, we show that the inner
slope of the dark matter halo before the baryons collapsed to form the
lens galaxy is $\gamma_{i}<1.4$ (68\% CL), only marginally consistent
with the highest-resolution cold dark matter simulations that indicate
$\gamma_{i}$$\sim$1.5. This might indicate that either adiabatic
contraction is a poor description of early-type galaxy formation or
that additional processes play a role as well. Indeed, the apparently
isothermal density distribution inside the Einstein radius, is not a
natural outcome of adiabatic contraction models, where it appears to
be a mere coincidence. By contrast, we argue that isothermality might
be the result of a stronger coupling between luminous and dark-matter,
possibly the result of (incomplete) violent relaxation processes
during the formation of the innermost regions of the galaxy. Hence, we
conclude that galaxy D appears already relaxed $\sim$8~Gyr ago. We
briefly discuss the importance of our results for lens statistics and
the determination of the Hubble Constant from gravitational-lens time
delays.
\end{abstract}


\section{Introduction}

Understanding the internal structure of galaxies, is crucial to
explain the variety of their observed morphologies and the physical
processes that created them. It also provides an opportunity to test
physics on galactic scales.

Because early-type (E/S0) galaxies are relatively simple systems,
dynamical modeling of their internal structure can be attempted with
hope for success (e.g. de Zeeuw \& Franx 1991; Bertin \& Stiavelli
1993; Merritt 1999). Knowledge of this internal structure provides a
vital discriminant to the processes involved in their formation
(e.g. van Albada 1982), which is currently a highly controversial
issue (e.g. Schade et al.\ 1999).

In particular, in the cold dark matter (CDM) cosmological model,
galaxies are embedded in dark matter halos, which dominate the total
mass of the system and have a characteristic mass distribution
(Navarro, Frenk \& White 1997, hereafter NFW; Moore et al.\ 1998;
Bullock et al.\ 2001; Ghigna et al.\ 2000). Measuring the distribution
of dark matter in E/S0 galaxies would therefore not only shed light on
their formation history, but also provide a powerful test of the CDM
cosmological model. This test is similar, but complementary in the
scales and physical conditions probed, to the still controversial
tests performed using the kinematics of low surface-brightness and
dwarf galaxies (e.g. McGaugh \& de Blok 1998; van den Bosch et al.\
2000; Swaters, Madore \& Trewhella 2000; Salucci \& Burkert 2000;
Borriello \& Salucci 2001; van den Bosch \& Swaters 2001; de Blok et
al.\ 2001; de Blok \& Bosma 2002; Jimenez, Verde \& Peng 2002).

Unfortunately, E/S0 galaxies generally lack simple dynamical tracers
at large radii, such as HI in spiral galaxies, and little is known
about their kinematics beyond the region dominated by stellar mass
(1--2 effective radii, R$_e$). Hence, the amount and distribution of
dark matter within E/S0 galaxies remain to date poorly constrained
even in the local Universe (e.~g. Bertin et al.\ 1994; Franx, van Gorkom \&
de Zeeuw 1994; Carollo et al.\ 1995; Rix et al.\ 1997; Gerhard et al.\
2001). In a few exceptional cases kinematic tracers at large radii
have been found and used to investigate the presence and distribution
of dark matter halos (e.~g. Mould et al.\ 1990; Franx et al.\
1994; Hui et al.\ 1995; Arnaboldi et al.\ 1996). However,
these methods do not seem likely to be applicable at significant
look-back times and thus to the investigation of the evolution of 
the internal structure of E/S0 galaxies.

An independent measure of the mass distribution of E/S0 galaxies is
provided by gravitational lensing, which is not sensitive to its
dynamical state. For example, gravitational lensing alone has been
used to show the existence of dark matter in individual E/S0 galaxies
(e.~g. Kochanek 1995) and recently to constrain the dark matter halo
profile from gravitational-lens statistics (e.g. Keeton 2001; Kochanek
\& White 2001), and galaxy-galaxy lensing signal (e.g. Seljak
2002). The combination of dynamical and lensing analyses can
significantly reduce the degeneracies inherent to each method and
result in tighter constraints on the dark-matter distribution, as well
as on the luminous and total mass distribution. In addition, most
lenses are found at intermediate redshift ($0.1<z<1$) and therefore
offer the possibility to probe the mass distribution at significant
look-back times.

\subsection{The Lenses Structure and Dynamics (LSD) Survey}

To investigate the internal structure of E/S0 galaxies and test CDM
predictions, we have started a program to measure the kinematic
profiles of a sample of E/S0 galaxies that are also gravitational
lenses; the {\sl Lenses Structure and Dynamics Survey} (hereafter the
LSD Survey \footnote[2]{see `www.astro.caltech.edu/$\sim$tt/LSD' or
`www.its.caltech.edu/$\sim$koopmans/LSD'}). By means of deep
spectroscopy, using the Echelle Spectrograph and Imager (ESI) at the
W.~M.~Keck--II Telescope, we are deriving extended kinematic profiles
(beyond the effective radius) for the nearest and brightest lenses,
along major and sometimes minor axes, while for the most distant and
faint ones we are measuring a luminosity-weighted velocity dispersion.
The ultimate aim of the LSD Survey is to produce a database of
measurements of internal kinematics for the largest possible number of
E/S0 lens galaxies, covering a range of redshifts and masses.

The target lenses have to: {(i)} have E/S0 morphology, as
determined from Hubble Space Telescope (HST) images; {(ii)} be
relatively undisturbed by the light of the lensed object at ground
based resolution; {(iii)} be relatively isolated, since a nearby
cluster, large group, or massive companion, would introduce additional
uncertainties in the measurement of the mass of the lens; {(iv)} be
bright enough that accurate internal kinematics can be determined in a
few hours of integration. A preference is given to objects where the
redshift of the background source is known. If the redshift of the
source is not known and the source is bright enough that the redshift
can be determined in a few hours of integration time, the redshift is
also determined.

In addition to help clarifying the origin of E/S0 galaxies, other
important results will be obtained if a better understanding of the
internal structure of E/S0 and its evolution with redshift is
achieved. For example we can mention the determination of the Hubble
constant from time-delay, and the calculations of lens cross-sections.

\subsection{A case study: MG2016+112}

In a previous paper (Koopmans \& Treu 2002; hereafter KT02) we
described the Keck and HST observations of the E/S0 lens galaxy D in
MG2016+112, the most distant spectroscopically confirmed lens galaxy
known ($z=1.004$).  A review of the current observations of this
system can be found in Koopmans et al.\ (2002; hereafter K02), who
also propose a detailed lens model, where the image separation is
caused predominantly by galaxy D, and find a central velocity
dispersion $\sigma$=320--340 \kms\ for an isothermal total mass
distribution. This value should correspond closely to the central
stellar velocity dispersion (e.~g. Kochanek 1994; Kochanek et al.\
2000).  Our direct measurement ($\sigma=328\pm32$ \kms; KT02) shows
that D is indeed a massive old E/S0 galaxy, possibly embedded in a
proto-cluster, with a central stellar velocity dispersion in agreement
with that inferred from the lens models.

In this paper we present a detailed analysis of the internal structure
of galaxy D, based on the observed velocity dispersion and the
exquisite HST images available from the CASTLES database. In
particular, we focus on the density profiles of (i) the dark matter
halo and (ii) the total luminous plus dark mass distribution.  Apart
for its high redshift (corresponding to $\sim 8$ Gyr look-back
time\footnote[3]{throughout the paper, we assume the Hubble Constant,
the matter density, and the cosmological constant to be
H$_0$=65~\kms\,Mpc$^{-1}$, $\Omega_{\rm m}=0.3$ and
$\Omega_{\Lambda}=0.7$, respectively.}), this system is particularly
interesting because its Einstein radius (R$_{\rm
Einst}$=13.7$\pm$0.1\,kpc) is much larger than the effective radius of
the lens galaxy ($R_{\rm e}=2.7\pm0.5$\,kpc). For this reason, we can
use gravitational lensing to probe the mass distribution to much
larger radii than is typically possible, even in local E/S0 galaxies.

The paper is organized as follows. First, the luminous and dark-matter
models are described in Sect.~2. In Sect.~3, constraints are placed on
the dark and luminous matter distributions inside the Einstein radius,
and compared to CDM predictions in Sect.~4. In Sect.~5, we use the
results found for MG2016+112 to predict observables for other similar
systems. Conclusions and a discussion are given in Sect.~6. Throughout
this paper, $r$ is the radial coordinate in 3-D space, while $R$ is the
radial coordinate in 2-D projected space.

\section{Models}

Based on the lens models in K02, one finds a total mass enclosed
within the Einstein radius, $M(<R_{\rm Einst})\equiv
M_E=1.1\times10^{12}$ M$_{\odot}$. The enclosed mass has been
corrected for the slight ellipticity of the lens potential and,
because the source lies almost on the observer-lens axis, is nearly
independent of the mass distribution assumed for the lens model
(e.g. Schneider, Ehlers \& Falco 1992; Chapter 8). An error of
$\sim10\%$ is assumed for the enclosed mass (K02). This error is
larger than what one might expect from a range of lens models
(e.g. see Kochanek 1991), but is chosen to include any remaining
uncertainties related to possibly undetected mass contributions from
nearby galaxies. Note, however, that to first order, the contribution
from the most massive nearby structures has been corrected for in the
lens models (K02). The error estimate on the enclosed mass can thus be
regarded as conservative. Later, we will estimate the effect of this
potential systematic error on our conclusions and show it to be
negligible.

The total luminosity of the system is $(1.6\pm0.2)
10^{11}$~L$_{\odot,B}$ (KT02), implying an average mass-to-light ratio
M/L$_B=8\pm1$~M$_{\odot}$/L$_{\odot,B}$ inside the Einstein radius,
assuming an \dv surface-brightness profile.  This value is similar to the
values found for the stellar populations of local E/S0 galaxies
(e.g. Gerhard et al.\ 2001). However, considering the much smaller age
of the stellar populations of galaxy D -- estimated to be $3.0\pm0.8$
Gyr and not larger than the age of the Universe at $z=1.004$ ($\sim6$
Gyr; KT02) -- the stellar M/L ratio of galaxy D must be smaller than
in the local Universe, due to luminosity evolution of the stellar
population. Given the large value of M/L, we conclude that a
significant amount of dark matter is likely to be present inside
$R_{\rm Einst}$. 

We choose to describe the mass distribution of
MG2016+112 with two spherical components, one for the luminous matter
and one for the dark matter halo. In particular, we model the luminous
mass distribution with a Hernquist (1990) model
\begin{equation}
\rho_L(r)=\frac{M_* r_*}{2\pi r(r+r_*)^3}, 
\label{eq:HQ}
\end{equation}
that well reproduces the \dv surface brightness profile for $r_* = R_e
/ 1.8153$ ($M_*$ is the total stellar mass) and is simple to treat
analytically (e.g. Ciotti, Lanzoni \& Renzini 1996). The dark-matter
distribution is modeled as
\begin{equation}
\rho_d(r)=\frac{\rho_{d,0}}{(r/r_b)^{\gamma}(1+(r/r_b)^2)^{\frac{3-\gamma}{2}}}
\label{eq:DM}
\end{equation}
which closely describes a NFW profile for $\gamma=1$, and has the
typical asymptotic behavior at large radii found from numerical
simulations of dark matter halos $\propto r^{-3}$ (e.~g. Ghigna et
al.\ 2000). Note that, for $r_b$ significantly exceeding $R_{\rm
Einst}$, the particular choice of the outer slope has only a
negligible effect on the kinematics of the lens galaxy. In
particular, for $r_b\rightarrow\infty$ and $\gamma=2$, Eq.~\ref{eq:DM}
reduces to a singular isothermal sphere, a functional form often used
to describe dark matter halos. In addition, Eq.~\ref{eq:DM} provides a
simple parametrization of the controversial inner slope for $r<r_b$,
the region better constrained by our observations. The length scale
($r_b$) and the density scale ($\rho_{d,0}$) determine the virial mass
of the dark matter halo (e.~g. Bullock et al.\ 2001).

The velocity dispersion is computed solving the spherical Jeans
equation (e.g. Binney \& Tremaine 1987), assuming an Osipkov-Merritt
(Osipkov 1979; Merritt 1985a,b) parametrization of the anisotropy
$\beta$:
\begin{equation}
\beta(r)=1-\frac{\sigma^2_{\theta}}{\sigma_{r}^2}=\frac{r^2}{r^2+r^2_i},
\label{eq:OM}
\end{equation}
where $\sigma_{\theta}$ and $\sigma_{r}$ are the tangential and radial
component of the velocity dispersion.  Note that $\beta>0$ by
definition and therefore tangentially anisotropic models are not
considered in this analysis.  Finally, the velocity dispersion as
observed through our spectroscopic aperture ($\equiv \sigma_{ap}$;
effectively corresponding to a circular aperture of radius
$0\farcs65\sim 2 R_e$) is computed as the average of the projected
velocity dispersion weighted by the surface brightness.

\section{Luminous and dark matter in MG2016+112}

\subsection{Constraints from kinematics, lensing geometry, and surface brightness distribution}

The effective radius and the total mass ($M_E$) within the Einstein radius 
are fixed by the observations (KT02). This leaves four free parameters
in the model: the inner slope ($\gamma$) of the dark matter halo, the
length scale of the dark matter component ($r_b$), the mass-to-light
ratio of the luminous component (M$_*/L_B$) and the anisotropy radius
($r_i$). In Fig.~\ref{fig:mod2D} we show the likelihood
contours\footnote[4]{Gaussian error distributions are assumed.} of
$\gamma$ and M$_*/L_B$, given the observed velocity dispersion
($\sigma_{ap}=304\pm27$ \kms) for a range of values of $r_b$. Since
changing $r_i$ has very little effect on the likelihood contours, only
the ones obtained for $r_i=R_e$ are shown (for more discussions see also 
also the rest of this section and in Sect.\ref{sec:prediction}).
\begin{inlinefigure}
\begin{center}
\resizebox{\textwidth}{!}{\includegraphics{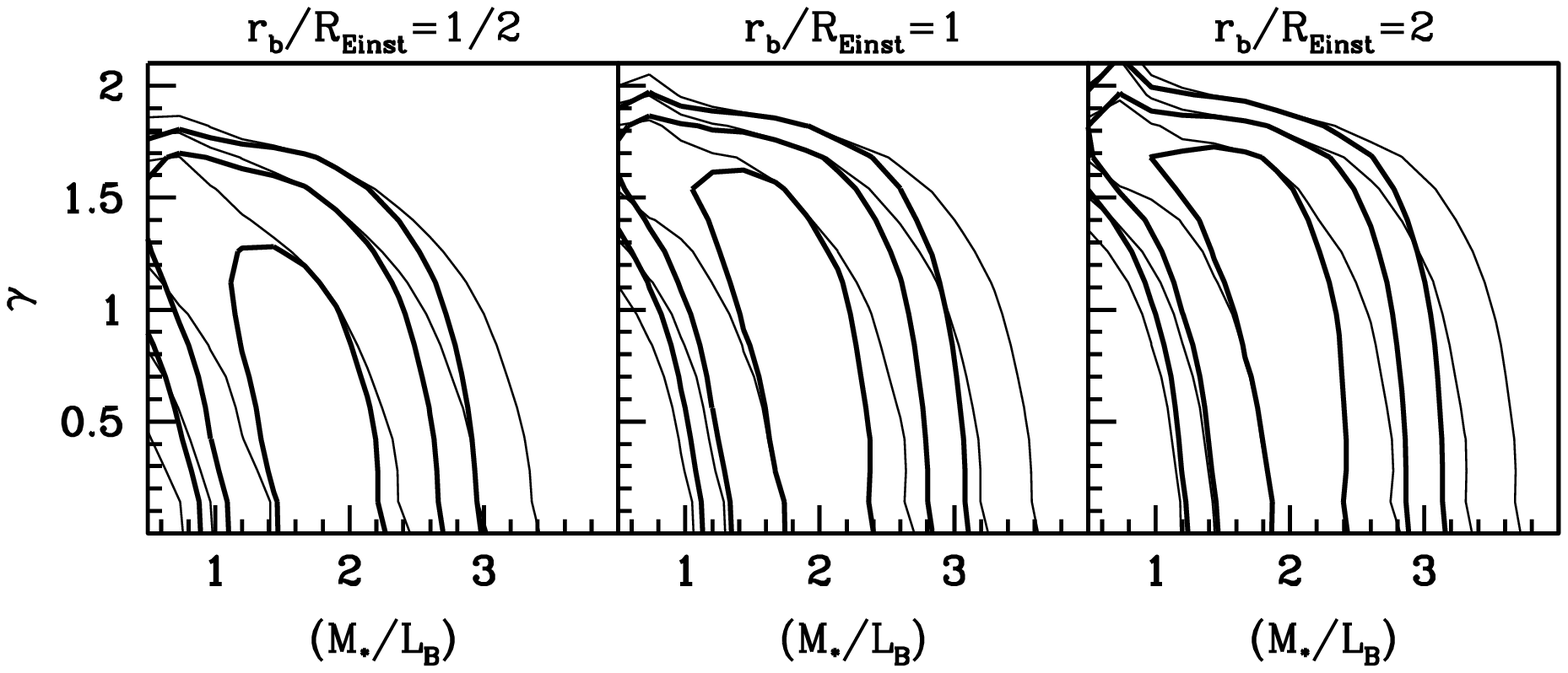}}
\end{center}
\figcaption{Constraints on the inner slope of the dark matter halo
($\gamma$) and the stellar mass-to-light ratio ($M_*/L_B$, in solar
units). Thin contours represent the 68\%, 95\%, 99\% confidence limits
given M$_E$, R$_{\rm e}$ and $\sigma_{ap}$. The thick contours
represent the same limits, if the independent measurement of $M_*/L_B$
from the evolution of the FP is included (Eq.~\ref{eq:FPev}). The
subpanels present the results for different values of the dark-matter
length scale ($r_b$). The contours change negligibly for different
values of the anisotropy radius anisotropy radius ($r_i/R_e=0.5,1,2$),
thus only the case $r_i/R_e=1$ is shown. See Sect.~3 for a detailed
discussion. \label{fig:mod2D}}
\end{inlinefigure}
Two main conclusions can be drawn from Fig.~1. First, since M$_*/L_B
\la4$\,\mlsun (99\% CL), i.e. smaller than the total
$M/L_B$=$8\pm1$\,\mlsun, dark matter is present within $R_{\rm Einst}$
and contributes $\ga 50\%$ (99\% CL) of the mass. As an independent
check, if we assume that $M_E$ arises from the luminous component only
with a constant M$_*/L_B=8$\mlsun, we find $\sigma_{ap}=530$~\kms\
(virtually independent of the isotropy radius), inconsistent with the
observed value at better than 8$\sigma$. Second, we find that for
large values of $r_b/R_{\rm Einst}$ (e.~g. $r_b/R_{\rm Einst}=2$ shown
in Fig.~1) and small values of $M_*/L_B$, $\gamma$ approaches the
value 2. Indeed, results from calculations with M$_*/L_B\rightarrow 0$
and $r_b\rightarrow \infty$, show that $\gamma=2.0\pm0.1$ (1--$\sigma$
error).  In other words, the total mass can be very well described by
a single power-law density distribution $\rho_t\propto r^{-\gamma'}$
with effective\footnote[5]{We use the term effective slope to
emphasize that this is the slope that we get with a single power
law. The real mass distribution could change slope between the
constraints given within the spectroscopic aperture radius and at
$R_{Einst}$, as long as the effective trend is preserved.}  slope
$\gamma'=2.0\pm 0.1$. A systematic error of 10\% on $M_{\rm E}$
changes the inferred slope by only 0.05, whereas a 20\% error
(1--$\sigma$) on $R_{\rm e}$ affects the result by only
0.02. Similarly, a 20\% error (1--$\sigma$) on the assumed aperture
radius of $0\farcs65$ changes the result by 0.03. Even a factor 10
deviation from $r_i=R_{\rm e}$ results in $\la 0.05$ error, although
the effect on the central velocity dispersion\footnote[6]{defined as
the luminosity-weighted velocity dispersion within a circular aperture
of radius $R_{\rm e}/8$} $\sigma$ in that case is much more profound,
with significantly larger values of $\sigma$ for smaller $r_i$
(i.e. more radial orbits, see Fig.~4).  Hence, $\gamma'$ is robust
against reasonable changes in the parameters $r_i$, $R_{\rm e}$,
$M_{\rm E}$, the aperture radius, and $r_b\ga R_{\rm Einst}$.

We can make a very simple argument to understand why the effective
slope is so tightly constrained by our observations. For a total mass
profile $\rho_{\gamma'}\propto r^{-\gamma'}$ the enclosed mass is
$M(<r)\propto r^{3-\gamma'}$. Therefore, the mass within the effective
radius ($M_{e}$) depends on the mass within the Einstein radius
($M_E$) as
\begin{equation}
M_e=M_E \left(\frac{R_{e}}{R_{Einst}}\right)^{3-\gamma'} \equiv
M_E x_E^{\gamma'-3},
\label{eq:MM}
\end{equation}
where $x_E\equiv R_{Einst} / R_e$ has been introduced for simplicity
of notation. Hence, assuming $M_e \propto \sigma^2$ for a given $R_e$,
\begin{equation}
\delta \gamma' \approx \frac{2 \delta \sigma}{\sigma |\log (x_E)|},
\label{eq:errors}
\end{equation}
where $\delta \gamma'$ and $\delta \sigma$ are the errors on the
effective slope and luminosity-weighted 
velocity dispersion, respectively. Using the
values for MG2016+112 in Eq.~\ref{eq:errors} we recover $\delta
\gamma'\approx 0.1$ in good agreement with the error inferred from our
model.

\subsection{Further constraints: the evolution of the Fundamental Plane}

Recent studies have shown that E/S0 galaxies both in clusters and in
the field define a tight Fundamental Plane (FP; Dressler et al.\ 1987;
Djorgovski \& Davis 1987) out to $z\sim0.7-0.8$ (van Dokkum et al.\
1998; Treu et al.\ 1999, 2002; see also Kochanek et al.\ 2000), with
slopes very similar to the ones observed in the local
Universe. Assuming that galaxy D in MG2016+112 lies on a FP with
slopes as in the local Universe, we can obtain an additional
constraint on its internal structure. In fact, the evolution of the
intercept of the FP with redshift can be related to the evolution of
the average effective mass to light ratio ($\Delta \log M/L_B$; see
e.~g. Treu et al.\ 2001 for discussion). For D we obtained $\Delta
\log M/L_B=-0.62\pm0.08$, i.e. intermediate between the cluster and
field value (KT02). We can use this measurement to infer $M_*/L_B$ of
D assuming that
\begin{equation}
\log (M_*/L_B)_{z}= \log (M_*/L_B)_{0} + \Delta \log (M/L_B),
\label{eq:FPev}
\end{equation}
where the second term on the right hand side of the equation is
measured from the evolution of the FP.  The first term on the right
hand side of the equation can be measured for local E/S0 galaxies: for
example using the data from Gerhard et al.\ (2001; and references
therein), we find a good correlation between $\sigma^2 R_e$ and
$M_*/L_B$. Using the values of MG2016+112, this correlation yields
$M_*/L_B(z=0)=7.3\pm2.1$\,\mlsun, where the error includes the
contribution from the scatter between dynamical and stellar population
measurements\footnote[7]{A very similar result is obtained by taking
simply the average stellar M/L of the sample
$7.8\pm2.7$\mlsun.}. Hence, from Eq.~\ref{eq:FPev} we find
$M_*/L_B(z=1.004)=1.8\pm0.7$\,\mlsun.  We note two things: {(i)} the
value $M_*/L_B=8$\,\mlsun\ required to explain the enclosed mass solely
by luminous matter is inconsistent with the above,
independently--derived, value at the 9--$\sigma$ level, and {(ii)} the
allowed range of $M_*/L_B$ found from our dynamical modeling (Fig.~1)
overlaps very nicely with the value derived from the evolution of the
FP.  Using this additional constraint we obtain the probability
contours plotted as thick solid lines in Fig.~\ref{fig:mod2D}. Again, the
results depend very modestly on the anisotropy radius, while the slope
$\gamma$ becomes steeper as $r_b$ increases. In general, $\gamma=2$
can be ruled out at better than 2--4 $\sigma$ (depending on $r_b$),
and therefore the dark matter profile is {\sl not} isothermal.
\begin{inlinefigure}
\begin{center}
\resizebox{\textwidth}{!}{\includegraphics{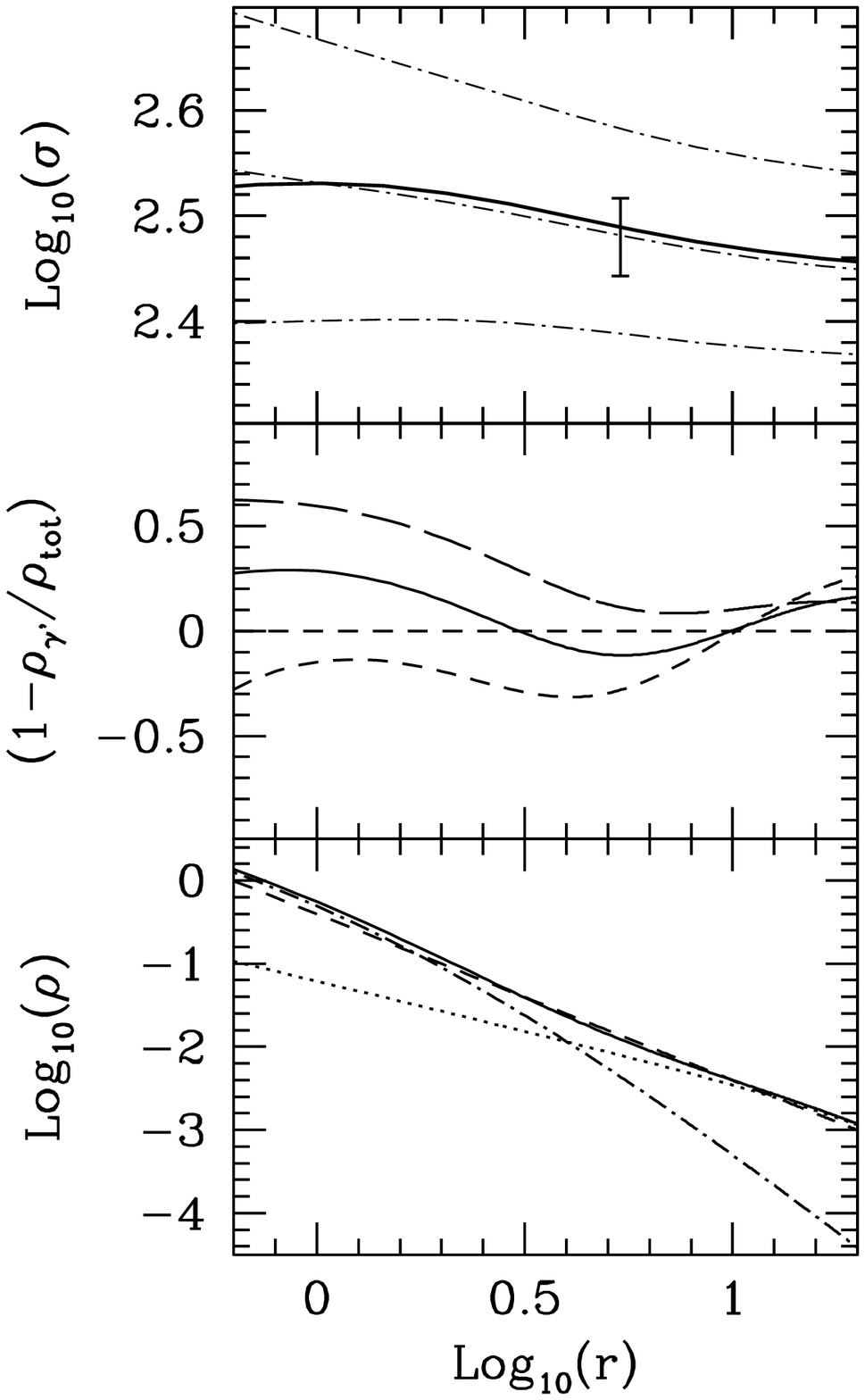}}
\end{center}
\figcaption{Dark and luminous matter best-fit density (in units of
$10^{10}$~M$_{\odot}\,$kpc$^{-3}$) and velocity dispersion profiles
(in \kms) as function of radius (in kpc). Lower panel: Density profile
of the luminous component (dot-dashed), of the dark matter halo
(dotted) and the sum of the two (solid) for several characteristic
best-fitting models.  An $r^{-2}$ density profile (dashed) reproduces
the total density profile remarkably accurately. Middle panel:
Relative residuals of the total mass distribution ($\rho_{tot}$) with
respect to power laws $\rho_{\gamma'}=r^{-\gamma'}$, with
$\gamma'=2.2,2,1.8$ from top to bottom.  Upper panel:
Luminosity-weighted velocity dispersion profile as a function of
circular aperture. The solid lines are obtained for the best-fitting
two-component model. The thin dot-dashed lines correspond to power-law
mass distributions with $\gamma'=$2.2,2,1.8 from top to bottom. The
error bar represents our measurement $\sigma_{ap}=304\pm27$\kms. We
assume $M_*/L_B=1.8$\mlsun.
\label{fig:density}}
\end{inlinefigure}

Let us now examine the total mass distribution of the two-component
model in more detail with the help of the quantities plotted in
Fig.~2. First (lower panel), we show the density profiles of a
characteristic best-fitting model ($r_b=2R_e$, $r_i=R_e$;
$M_*/L=1.8$\,\mlsun). Even though luminous and dark matter themselves
are not isothermally distributed, the total mass distribution is very
close to isothermal, confirming what we found in Section~3.1. Second
(middle panel), we show the relative difference between a power law
mass profile $\rho_{\gamma'}\propto r^{-\gamma'}$ and the total mass
profile obtained with our two-component model. A profile with
$\gamma'=2$ describes the total mass profile very well. By contrast,
$\gamma'=1.8$ or $\gamma'=2.2$ result in significantly worst
representations of the total mass profile. Third (upper panel), we
compare the luminosity-weighted velocity dispersion profile of our
two-component model with those obtained for $\gamma'=$1.8, 2.0 and
2.2. The $\gamma'=2$ power law mass profile and the two-component
model produce almost identical velocity dispersion profiles, while
$\gamma'=1.8,2.2$ are cleary inconsistent with the observed
$\sigma_{\rm ap}$ (point with error bar in the plot).  For
completeness, we note that we find $\sigma/\sigma_{ap}=1.06$,
consistent with our assumption of $1.08\pm0.05$ in KT02.

Finally, we check that our best-fitting models correspond indeed to
physical solutions in the sense that they can be generated by a
positive distribution function (i.e. they are {\it consistent} models,
see e.g. Ciotti 1999 and references therein for discussion).  For our
choice of values of $r_i/R_e$, a luminous Hernquist model embedded in
a total isothermal mass distribution always satisfies the sufficient
condition given in Eq.~6 of Ciotti (1999), and therefore our
best-fitting models are {\it consistent}.

\section{Comparison with CDM models}

In order to make a proper comparison with CDM predictions, we have to
consider that the universal profiles of dark matter halos can be
significantly modified by the formation of a galaxy within the halo. A
simple treatment of this phenomenon can be given by assuming the
so-called ``adiabatic approximation'' (Blumenthal et al.\ 1986; see
also Mo, Mao \& White 1998). Under this approximation, for any given
initial mass distribution (baryons are assumed to follow dark matter
at the beginning) we can calculate the final dark matter distribution
after the galaxy has collapsed, for example to an Hernquist mass
distribution (e.~g. Keeton 2001; see also Rix et al.\ 1997). Here, we
use the approximation backwards to recover the initial dark matter
density profile. In order to facilitate comparison to CDM predictions,
we fit the same functional form as in Eq.~\ref{eq:DM} to the initial
density profile by minimizing the fractional difference in density
integrated over 0--100 kpc, and derive the initial length scale
$r_{b,i}$ and an initial slope $\gamma_i$. Although it was not
guaranteed {\it a priori}, we find that this form gives a reasonable
description of the initial density profile as well.

Before proceeding to the calculation, we reduce the degrees of freedom
of the problem by noticing that CDM models indicate that the length
scale of systems as massive as galaxy D should be significantly larger
than $\sim 15$ kpc, even at $z\sim1$ (Bullock et al.\ 2001).
Therefore, in the spirit of the comparison with CDM predictions, we
can limit our parameter space to $r_b>R_{\rm Einst}$, where the
results depend very little on the exact value of $r_b$. The length
scale does not change significantly ($<30$\% depending on the slope
for $r_{bi}<40$~kpc) during collapse, therefore our assumption on
$r_b$ being large because $r_{b,i}$ is large is justified in this
context. For definiteness, we pick $r_b=2R_{\rm Einst}$ so that the
observed profiles are shown in Fig.~\ref{fig:mod2D}
and~\ref{fig:density}.  With this assumption, the initial slope
$\gamma_i$ depends only on the observed slope $\gamma$ and on
$M_*/L_B$.

In Fig~\ref{fig:adiabatic} we show the contour levels of $\gamma_i$ in
the $\gamma-M_*/L_B$ plane. As $M_*/L_B\rightarrow0$ , the effect of
baryons becomes negligible and $\gamma\rightarrow\gamma_i$. On the
other hand, as $M_*/L_B$ increases, the collapse becomes increasingly
important and flatter initial dark matter halos correspond to
increasingly steeper final halos.  To constrain $\gamma_i$ with the
observations, we overlay the likelihood contours of $\gamma-M_*/L_B$
derived from our dynamical model (Fig.~1), and consider only the
portion of the $\gamma_i$ contours enclosed by the likelihood
contours. Considering only monotonically decreasing profiles, the
range $\gamma_i=0.0-1.4$ maps onto a very tight range of final slopes
$\gamma\approx 1.30-1.75$. In conclusion -- within the context of this
simple adiabatic contraction model -- $\gamma_i=1.5$ is only
marginally consistent with our observations, whereas $\gamma_i=1$ fits
within the 68\% likelihood contour.
\begin{inlinefigure}
\begin{center}
\resizebox{\textwidth}{!}{\includegraphics{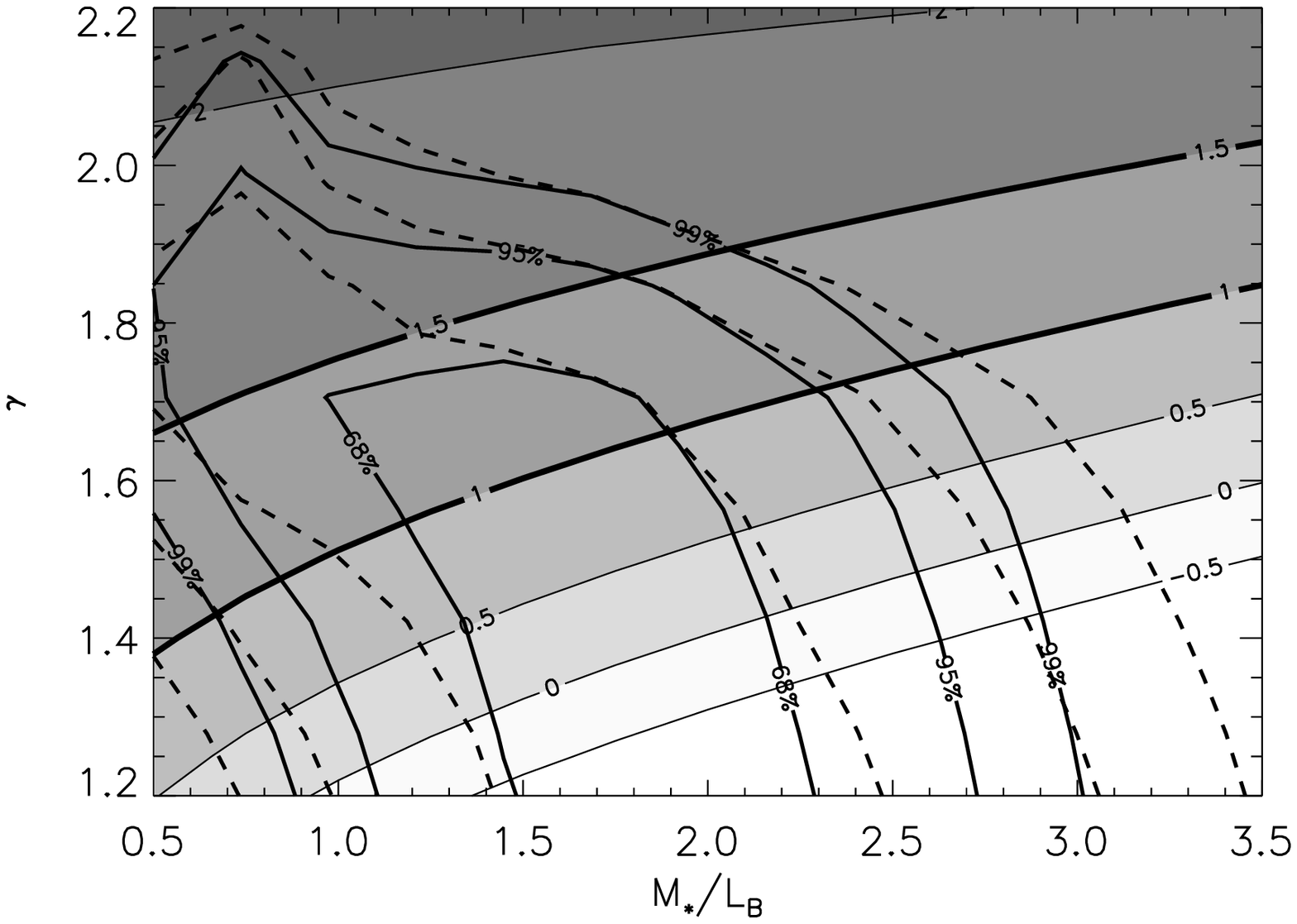}}
\end{center}
\figcaption{Relation between the final ($\gamma$)
and the initial ($\gamma_i$) inner slope of the CDM dark-matter halo
of galaxy D in MG2016+112, and the stellar mass-to-light ratio
($M_*/L_B$), based on a simple adiabatic contraction model (see text).
The loci corresponding to different initial slopes are shown, labeled
by $\gamma_i$. In particular, the two thick contours labeled 1.0 and
1.5 (=$\gamma_i$) indicate the NFW and Moore et al.\ (1998) profiles,
respectively.  The probability contours given the observations of
MG2016+112 for a model with $r_{\rm d}=2\,R_{\rm Einst}$ and
$r_i=R_{\rm e}$ (see Fig.1) are overplotted and labeled by the
probability level. As in Fig.~1 solid lines include the constraint
from the FP evolution, dashed lines do not.
\label{fig:adiabatic}}
\end{inlinefigure}

\section{Kinematics of Hernquist luminosity density profiles
in a logarithmic gravitational potential}

\label{sec:prediction}

In Sect.~3 we have shown that the total mass distribution of galaxy D
is well described by an isothermal sphere in the range $\sim$ 1--15
kpc. Here, we generalize the results found for galaxy D to compute the
velocity dispersion profiles that we expect to find for other lens
E/S0 galaxies.

Let us assume that E/S0 galaxies have total mass profiles that follow
$\rho_{tot}\propto r^{-2}$ between a fraction of an effective radius
and a few effective radii.  Solving the Jeans Equation for a Hernquist
luminosity-density profile embedded in a logarithmic potential, we can
compute the velocity dispersion profile, depending only on the
effective radius ($R_e$), the ``Einstein velocity dispersion''
($\sigma_{\rm Einst}\equiv\sqrt{G\,M_{\rm E}/R_{\rm Einst}})$, and the
anisotropy radius ($r_i$) .  

In Fig.~\ref{fig:prediction}, we show the luminosity-weighted velocity
dispersion profile (in units of $R_e$ and $\sigma_{\rm Einst}$), for
several values of $r_i$.  Note that the profiles approach the same
curve for large $R/R_e$ independently of $r_i$. Hence, the effect of
changing $r_i$ on the luminosity-weighted velocity dispersion beyond
the effective radius, is not as dramatic as in the inner portions of
E/S0 galaxies.  For this very reason, the luminosity-weighted velocity
dispersion does not firmly constrain $r_i$. On the other hand, the
velocity dispersion profile itself is much more sensitive to
$r_i$. Therefore, a spatially resolved measurement of $\sigma(r)$,
extending from a fraction of to a few $R_e$, will allow us to
constrain both $r_i$ and the total mass distribution.
\begin{inlinefigure}
\begin{center}
\resizebox{\textwidth}{!}{\includegraphics{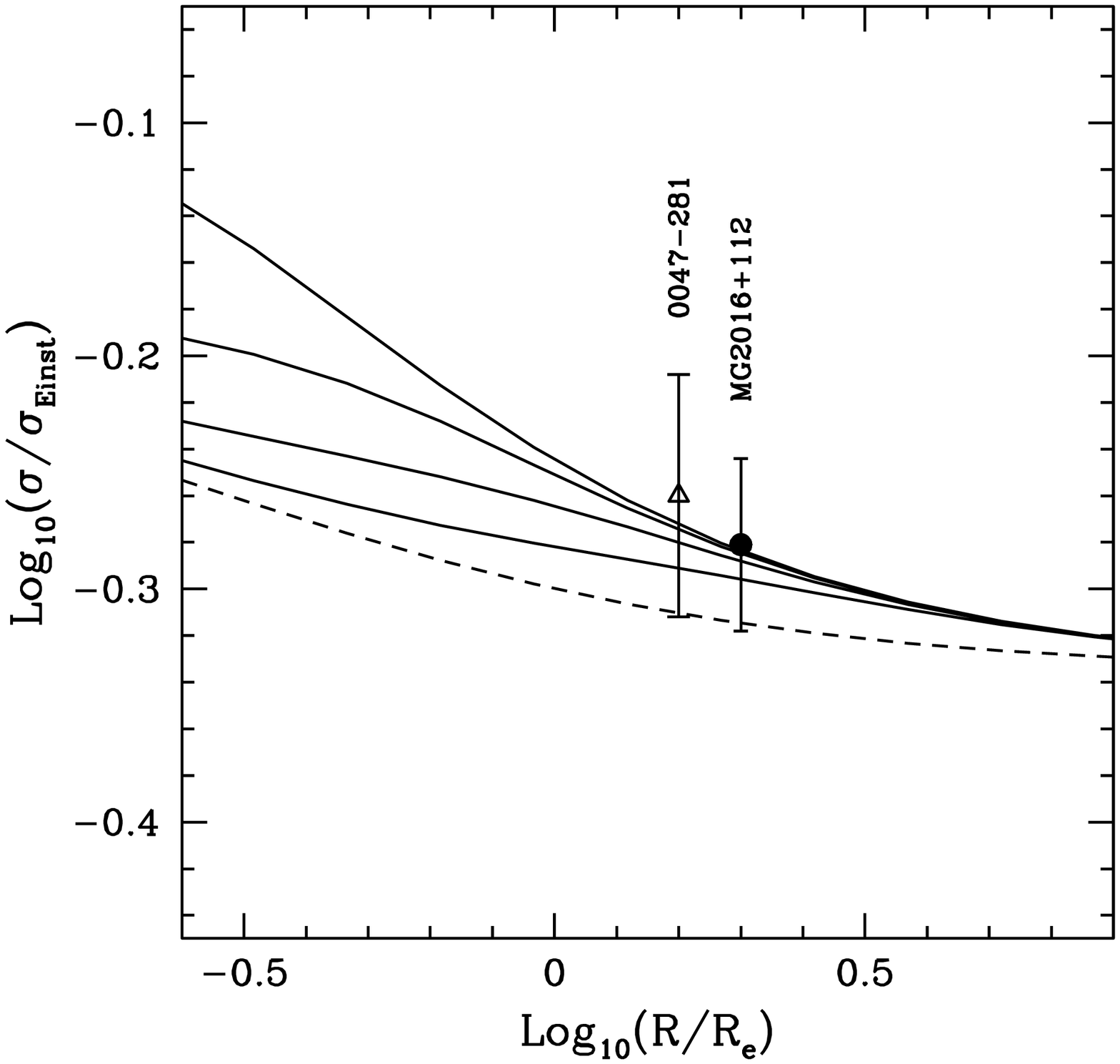}}
\end{center}
\figcaption{Luminosity-weighted velocity dispersion profile for a
Hernquist luminosity-density profile in logarithmic potential. The
axes are scaled in units of $R_e$ and $\sigma_{\rm Einst}$ (see
text). From top to bottom curves are shown for
$r_i/R_e$=$0.2$,$0.5$,1,2,$\infty$ (where $\infty$ corresponds to
isotropic models).  Note the smaller effect of changing anisotropy
radius $r_i$, as compared to changing the effective slope $\gamma'$ of
the total mass distribution (upper panels in Fig.~\ref{fig:density};
to facilitate comparison the two figures show the same range of
velocity dispersion and radius). Our measurement of MG2016+112 is
shown as a filled circle. The value for 0047-0281 using data from
Warren et al.\ (1998) and Kochanek et al.\ (2000) is shown as an open
triangle.
\label{fig:prediction}}
\end{inlinefigure}

To our knowledge, the only system (besides MG2016+112) satisfying the
LSD requirements\footnote[8]{Plus showing no evidence for rotation,
since this dynamical model is inappropriate for rotating systems.}
(Sec~1.1), for which velocity dispersion of the lens galaxy has been
measured is 0047-281 (Warren et al.\ 1996, 1998; the effective radius
and image separation are measured by Kochanek et al.\ 2000). As shown
in Figure~\ref{fig:prediction}, the observed velocity dispersion of
the lens galaxy in 0047-281 agrees very well with the value computed
with our model. In future papers we will present velocity dispersion
(profiles) for other systems, and further test this model.

\section{Summary \& Discussion}

\subsection{Summary}

We have shown that the combination of constraints from stellar
kinematics and gravitational lensing is extremely powerful in
determining the mass profiles of luminous and dark matter in E/S0
galaxies even at large cosmological distances. In particular, for galaxy D in
MG2016+112 ($z=1.004$) we have demonstrated that:

\noindent
{(i)} Dark matter is unambiguously present and accounts for
$>$50\% (99\% CL) of the mass within the Einstein radius. A mass
distribution following light with constant $M/L$ can be ruled out at
more than 8--$\sigma$ level. Considering an additional constraint on
$M_*/L_B$ from the evolution of the FP, the smallest allowable
fraction of dark matter increases to 60\% (99\% CL).

\noindent
{(ii)} The total mass distribution (dark+luminous) inside the Einstein
radius is well described by an isothermal sphere,
i.e. $\rho_{tot}\propto r^{-\gamma'}$ with $\gamma'=2.0\pm0.1$. This
result is robust and does not depend significantly on the dynamical
state of the system or on the modeling of the mass distribution (one
or two component models). The result is only weakly dependent on
errors on $M_E$, $r_i$, $R_{\rm e}$ and the aperture radius, which in
quadrature add at most another 0.1 to the error.

\noindent
{(iii)} Using two-component (dark and luminous) spherical models,
we find that the inner slope of the dark matter halo $\gamma<2.0$ at
the 95\% CL, i.~e. the halo is not isothermal inside the Einstein 
radius.

\noindent
{(iv)} Using a simple adiabatic contraction model, we relate the
observed inner slope $\gamma$ to the inner slope $\gamma_i$ of the
dark matter halo before collapse.  In particular we find the upper
limit $\gamma_i<1.4(2.0)$ at 68(99)\% CL, marginally consistent with
the high resolution CDM simulations that indicate
$\gamma_{i}$$\approx$1.5 (Ghigna et al.~2000; but see Power et al.\
2002 for a discussion of the related uncertainties).

\smallskip
These results are robust because the stellar velocity dispersion and
gravitational lensing provide two independent mass measurements at two
well-separated radii, the effective radius (2.7~kpc) and the Einstein
radius (13.7~kpc $\sim 5 R_{\rm e}$). This leads to a very
well-defined slope inside the Einstein radius, and to stringent limits
on the fraction of dark matter.

Higher signal to noise -- and possibly spatially resolved observations
-- are needed to improve these constraints. However, this is unlikely
to be feasible with the current generation of space or ground-based
telescopes and will probably need the Next Generation Space Telescope
or large ground-based telescopes with adaptive optics. Nevertheless,
when the measurement and analysis of the dozen systems targeted by the
LSD Survey will be completed, we expect to be able to obtain more
stringent constraints by combining the results from the individual
systems.

Finally, in Sect.~\ref{sec:prediction} we have provided a simple test
of whether an isothermal total mass distribution is a good description
of other lens galaxies. In particular, we compute the
luminosity-weighted velocity dispersion profile for a luminous
Hernquist component in a logarithmic potential, which depends very
mildly on the anisotropy radius, especially at radii larger than
$R_e$. The measured velocity dispersion of the only system available
besides MG2016+112 (0047+281) agrees very well with the curve computed
with this model. As new velocity dispersion measurements of E/S0 lens
galaxies become available, they can be readily compared to the
expectations of this simple model to test whether indeed the
isothermal mass profile is a general feature of the E/S0 population.

\subsection{Discussion}

In this paper and KT02 we presented the following simple picture of
the lens galaxy D in MG2016+112. The stellar populations are old and
metal rich. A significant amount of dark matter is present and the
mass distribution of the luminous and dark component are well
constrained, reproducing all the available observations.

However, the good agreement of the total mass profile with an $r^{-2}$
power law calls for a physical explanation. This seems not to be a
mere coincidence, given that a similar behavior has been observed in
some local ellipticals with kinematic tracers at extraordinary large
radii (e.~g. Mould et al. 1990; Franx et al.\ 1994; Rix et al.\ 1997).
Whether this galaxy formed by monolithic collapse, or by mergers of
smaller subunits in a hierarchical picture, why do these processes
conspire to produce an apparent isothermal mass distribution within
R$_{\rm Einst}$? In the case of adiabatic contraction, there appears
to be no reason why gas could not continue to contract or stop earlier
and give rise to a total mass distribution that is either steeper or
shallower than isothermal. Thus, formation by adiabatic contraction
seems not to be a satisfactory explanation, and it is worth seeking a
more fundamental explanation. By contrast, complete violent relaxation
produces isothermal distribution functions (Lynden-Bell 1967; Shu
1978), and would explain quite naturally the observations. However,
this only applies exactly for an infinite mass distribution. When
finite mass distributions are considered, (incomplete) violent
relaxation leads to an $R^{1/4}$ surface-brightness profile (e.g. van Albada
1982; Hjorth \& Madsen 1991, 1995), which goes as $r^{-2}$ within the
half mass radius, the region of interest here. For the two-component
case (luminous and dark), the process should be such as the two
components interact to produce the total $r^{-2}$ density profile,
while preserving the segregation of the luminous component in a
centrally concentrated density profile (see e.g. the two-component
models of Bertin, Saglia \& Stiavelli 1992; a similar outcome can be
produced by merging of galaxies with massive halos, see e.~g. Barnes
1992; Barnes \& Hernquist 1996).  In the CDM scenario, it is not clear
whether the chaotic changes in the gravitational potential during the
collapse of halos satisfy the conditions for violent relaxation
(e.g.~Flores \& Primack 1994 and references therein), although it
might not be inconceivable that violent relaxation of the inner parts
of E/S0 galaxies is close to complete. A complete discussion of these
issues goes beyond the aims of this paper. However, we would like to
emphasize that whatever physical processes are involved, the mass
distribution of galaxy D in MG2016+112 seems already relaxed $\sim 8$
Gyr ago.

Finally, we note that our poor knowledge of the slope of the mass
profile of lens galaxies is the major systematic uncertainty in the
determination of H$_0$ from time-delays (e.g. Koopmans \& Fassnacht
1999; Romanowsky \& Kochanek 1999; Koopmans 2001). The assumption of
isothermal mass distribution leads to
$H_0\sim60-70$~km\,s$^{-1}$\,Mpc$^{-1}$ (e.g. Koopmans \& Fassnacht
1999; Koopmans 2001), in good agreement with other independent
measurements. The result for MG2016+112 seems to indicate that the
systematic error -- the deviation from isothermality -- might indeed
be very small for E/S0 lens galaxies. However, the LSD Survey, by
measuring the average effective slope of the mass profile and its
scatter, will allow us to pin down the largest systematic uncertainty
in the global determination of H$_0$ from gravitational-lens time
delays.

\acknowledgments

We are grateful to E.~Agol, A.~Benson, G.~Bertin, R.~Blandford,
L.~Ciotti, R.~Ellis, M.~Stiavelli for the stimulating conversations
and insightful comments that helped in shaping this manuscript. Uros
Seljak is thanked for useful conversations. We thank the anonymous
referee for his prompt and careful report. We acknowledge financial
support from NSF and HST grants (AST--9900866; STScI--GO
06543.03--95A; STScI-AR-09222). Finally, the authors wish to recognize
and acknowledge the very significant cultural role and reverence that
the summit of Mauna Kea has always had within the indigenous Hawaiian
community.  We are most fortunate to have the opportunity to conduct
observations from this mountain.

\clearpage

\clearpage

\clearpage


\begin{thebibliography}{}

\bibitem[Arnaboldi et al. 1996]{Arna96} Arnaboldi, M. et al. 1996, \apj, 472, 145
\bibitem[Barnes 1992]{B92} Barnes, J. 1992, \apj, 393, 484
\bibitem[Barnes \& Hernquist 1996]{BL96} Barnes, J. \& Hernquist, L., 1996, \apj, 471, 115
\bibitem[Bertin, Saglia \& Stiavelli (1992)]{BSS92} Bertin, G., Saglia, R.~P., Stiavelli, M., 1992, 384, 423
\bibitem[Bertin et al. 1994]{B94} Bertin, G. et al.\ 1994, A\&A, 292, 381
\bibitem[Bertin \& Stiavelli 1993]{BS93} Bertin, G., \& Stiavelli, M., 1993, Rep. Prog. Phys, 56, 493
\bibitem[Binney \& Tremaine 1987]{BT87} Binney, J. \& Tremaine, S, 1987, Galactic Dynamics, Princeton University Press, Princeton
\bibitem[Borriello \& Salucci 2001]{BS01} Borriello, A., Salucci, P., 2001, \mnras, 232, 285
\bibitem[Blumenthal et al. 1986]{B86} Blumenthal, G.~R., Faber, S.~M., Flores, R., Primack, J.~R. 
\bibitem[Bullock et al.\ 2001]{B01} Bullock, J.~S., Kolatt T.~S., Sigad, Y., Somerville, R.~S., Kravtsov, A.~V., Klypin, A.~A., Primack, J.~R., \& Dekel, A., 2001, MNRAS, 321, 598
\bibitem[Carollo et al.\ 1995]{C95} Carollo, C.~M., de Zeeuw, P.~T., van der Marel, R.~P., Danziger, I.~J., Qian, E.~E., 1995, \apj, 441, L25
\bibitem[Ciotti et al.\ 1996]{CLR96} Ciotti, L., Lanzoni, B., Renzini, A., 1996, MNRAS, 282, 1
\bibitem[Ciotti (1999)]{C99} Ciotti, L., 1999, \apj, 520, 574
\bibitem[de Blok \& Bosma 2002]{DB02} de Blok, W.~J.~G. \& Bosma, A., 2002, A\&A, in press (astro-ph/0201276)
\bibitem[de Blok et al.\ 2001]{dB01} de Blok, W.~J.~G., McCaugh, S.~S., Bosma, A., \& Rubin, V.~C., 2001, \apj, 552, L23 
\bibitem[de Zeeuw \& Franx 1991]{dZF91} de Zeeuw, T., \& Franx, M., 1991, ARA\&A, 29, 239
\bibitem[Djorgovski \& Davis 1987]{DD87} Djorgovksi S.~G., Davis M., 1987, ApJ, 313, 59 
\bibitem[Dressler et al.\ 1987]{D87} Dressler, A., Lynden-Bell, D., Burstein, D., Davies, R.~L., Faber, S.~M., Terlevich, R, Wegner G. 1987, ApJ, 313, 42
\bibitem[Flores \& Primack 1994]{FP94} Flores, R.~A. \& Primack, J.~R. 1994, \apj, 427, L1
\bibitem[Franx et al. 1994]{F94} Franx, M., van Gorkom J.~H., \& de Zeeuw, P.T. 1994, \apj, 436, 642
\bibitem[Gerhard et al. 2001]{G01} Gerhard, O., Kronawitter, A., Saglia, R.~P., \& Bender, R., 2001, \aj, 121, 1936
\bibitem[Ghigna et al.\ 2000]{G00} Ghigna, S., Moore, B., Governato,
F., Lake, G., Quinn, T., Stadel, J., 2000, \apj, 544, 616
\bibitem[Hernquist 1990]{H90} Hernquist, L., 1990, \apj, 356, 359
\bibitem[Hjorth \& Madsen(1991)]{1991MNRAS.253..703H} Hjorth, J.~\& Madsen, J.\ 1991, \mnras, 253, 703
\bibitem[Hjorth \& Madsen(1995)]{1995ApJ...445...55H} Hjorth, J.~\& Madsen, J.\ 1995, \apj, 445, 55
\bibitem[Hui et al. 1995]{Hui95} Hui, X., Ford, H.~C., Freeman, K.~C., Dopita, M.~A., 1995, \apj, 449, 592
\bibitem[Jimenez, Verde \& Oh 2002]{JVO} Jimenez, R., Verde, L., Oh, S.~P. 2002, \mnras, submitted (astro-ph/0201352)
\bibitem[Keeton (2001)]{K01} Keeton, C.~R. 2001, ApJ, 561, 46
\bibitem[Kochanek (1991)]{K91} Kochanek, C.~S., 1991, \apj, 371, 289
\bibitem[Kochanek (1994)]{K94} Kochanek, C.~S., 1994, \apj, 436, 56
\bibitem[Kochanek (1995)]{K95} Kochanek, C.~S., 1995, \apj, 445, 559
\bibitem[Kochanek et al. (2000)]{Koch} Kochanek, C.~S. et al. 2000, ApJ, 543, 131
\bibitem[Kochanek \& White(2001)]{2001ApJ...559..531K} Kochanek, C.~S.~\& 
White, M.\ 2001, \apj, 559, 531
\bibitem[Koopmans 2001]{Ko01} Koopmans, L.~V.~E., 2001, PASA, 18, 179
\bibitem[Koopmans \& Fassnacht (1999)]{KF99} Koopmans, L.~V.~E. \& Fassnacht, C.~D., 1999, \apj, 527, 513
\bibitem[Koopmans et al.\ (2002)]{K02} Koopmans, L.~V.~E., Garrett, M.~A., Blandford, R.~D., Lawrence, C.~R., Patnaik, A.~R., Porcas, R.~W., 2001, \mnras, in press (K02)
\bibitem[Koopmans \& Treu (2002)]{KT02} Koopmans, L.~V.~E. \& Treu, T., 2002, \apj, 568, L5 (KT02)
\bibitem[Lynden-Bell 1967]{LB67} Lynden-Bell, D., 1967, MNRAS, 136, 101
\bibitem[McGaugh \& de Blok(1998)]{1998ApJ...499...41M} McGaugh, S.~S.~\& de Blok, W.~J.~G.\ 1998, \apj, 499, 41
\bibitem[Merritt 1985]{M85a} Merritt, D. 1985a, \aj, 90, 1027
\bibitem[Merritt 1985]{M85b} Merritt, D. 1985b, \mnras, 214, 25
\bibitem[Merritt 1999]{M99} Merritt, D., 1999, PASP, 111, 129 
\bibitem[Moore et al.\ 1998]{M98} Moore, B., Governato, F., Quinn, T., Stadel, J. \& Lake, G., 1998, \apj, 499, L5
\bibitem[Mould et al. 1990]{M90} Mould, J.~R., Oke, J.~B., de Zeeuw, P.~T., Nemec, J.~M., 1990, \aj, 99, 1823
\bibitem[Navarro et al.\ 1997]{NFW} Navarro, J, Frenk, C.~S., \& White S.~D.~M, 1997, \apj, 490, 493
\bibitem[Osipkov 1979]{O79} Osipkov L.~.P., 1979, Pis'ma Astron. Zh., 5, 77
\bibitem[Power et al. 2002]{P02} Power, C., Navarro, J.~F., Jenkins, A., Frenk C.~S., White, S.~D.~M., Springel, V., Stadel, J., Quinn, T., 2002, \mnras, submitted, astro-ph/0201544  
\bibitem[Romanowsky \& Kochanek 1999]{RK99} Romanowsky, A.~J. \& Kochanek, C.~S., 1999, \apj, 516, 18 
\bibitem[Rix et al.\ 1997]{R97} Rix, H.~W., de Zeeuw, P.~T, Cretton,
N., van der Marel, R.~P., \& Carollo, C.~M.  1997, \apj, 488, 702
\bibitem[Salucci \& Burkert 2000]{SB00} Salucci, P., Burkert, A., 2000, \apj, 537, L9
\bibitem[Schade et al.\ 1999]{CFRS-Es} Schade D. et al., 1999, ApJ, 525, 31
\bibitem[Schneider et al. 1992]{SEF92} Schneider, P., Ehlers, J. \&
Falco, E.~E. 1992, Gravitational Lenses, Springer-Verlag, Berlin. 
\bibitem[Seljak (2002)]{S02} Seljak, U., 2002, \mnras, in press, astro-ph/0201450
\bibitem[Shu 1978]{S78} Shu, F.~H. 1978, \apj, 225, 83  
\bibitem[Swaters, Madore, \& Trewhella(2000)]{2000ApJ...531L.107S} Swaters, R.~A., Madore, B.~F., \& Trewhella, M.\ 2000, \apjl, 531, L107
\bibitem[Treu et al.\ (2001)]{T01} Treu, T., Stiavelli, M., Bertin G., Casertano, C., \& M{\o}ller, P. 2001, \mnras, 326, 237
\bibitem[Treu et al.\ (1999)]{T99} Treu, T., Stiavelli, M., Casertano, C., M{\o}ller, P., \& Bertin G. 1999, \mnras, 308, 1307
\bibitem[Treu et al.\ (2002)]{T02} Treu, T., Stiavelli, M., Casertano, C., M{\o}ller, P., \& Bertin, G. 2002, \apj, 564, L13
\bibitem[van Albada 1982]{vA82} van Albada, T.S., 1982, MNRAS, 201, 939
\bibitem[van den Bosch, Robertson, Dalcanton, \& de Blok(2000)]{2000AJ....119.1579V} van den Bosch, F.~C., Robertson, B.~E., Dalcanton, J.~J., \& de Blok, W.~J.~G.\ 2000, \aj, 119, 1579
\bibitem[van den Bosch \& Swaters(2001)]{2001MNRAS.325.1017V} van den Bosch, F.~C.~\& Swaters, R.~A.\ 2001, \mnras, 325, 1017
\bibitem[van Dokkum et al.\ (1998)]{pgd98} van Dokkum, P.~G., Franx, M., Kelson D.~D. \& Illingworth G.~D., 1998, ApJ, 504, L17
\bibitem[Warren et al. (1996)]{W96} Warren, S.~J., Hewett, P.~C., Lewis, G.~F., M{\o}ller, P., Iovino, A., Shaver P.~A., 1996, MNRAS, 278, 139 
\bibitem[Warren et al. (1998)]{W98} Warren, S.~J., Iovino, A., Hewett, P.~C., Shaver P.~A., 1998, MNRAS, 299, 1215 

\end{thebibliography}
\end{document}